# Observation of a Mott quantum spin Hall phase in twisted $WSe_2$

Yifei Jin[1*], Yaqi Ma[1*], Aoqian Zhang[1*], Nan Zhang[1], Ulf Lampe[1], Ivana Wong[1], Kenji Watanabe[2], Takashi Taniguchi[2], Tze Kin Cheung[1], Junwei Liu[1,3], Ning Wang[1,4†] and Kaifei Kang[1,3†]

[1]Department of Physics, The Hong Kong University of Science and Technology, Hong Kong SAR, China

[2]National Institute for Materials Science, 1-1 Namiki, 305-0044 Tsukuba, Japan

[3]IAS Center for Quantum Matter, Hong Kong University of Science and Technology, Hong Kong SAR, China

[4]State Key Laboratory of Optical Quantum Materials, Hong Kong University of Science and Technology, Hong Kong SAR, China

*These authors contributed equally: Yifei Jin, Yaqi Ma and Aoqian Zhang.

Emails: phwang@ust.hk, kfkang@ust.hk

**Quantum spin Hall (QSH) insulators and Mott insulators are conventionally regarded as distinct insulating phases, arising from band topology and strong Coulomb interactions, respectively. Here, we report the observation of QSH edge transport in a magnetic-field-stabilized Mott insulating state at half filling of the second moiré band in a 2.29° twisted $WSe_2$ device. This state exhibits a resistance plateau identical to that of the single-particle QSH state at full filling of the first moiré valence band, indicating the same number of helical edge channels. Electrical transport measurements reveal nearly quantized resistance that is insensitive to vertical electric field, out-of-plane magnetic field, and temperature below 5 K. Pronounced nonlocal transport and strong negative in-plane magnetoconductance further support helical edge conduction, establishing robust edge transport in the strongly correlated regime. Temperature-dependent Hall measurements reveal a characteristic temperature scale of approximately 10 K, corresponding to an energy scale of about 1 meV. Our results demonstrate that spin-conserved QSH edge states can persist in a half-filled, strongly correlated insulating phase and under external magnetic field, opening a route toward interaction-resilient topological transport in moiré quantum materials.**

## Main

Quantum spin Hall (QSH) insulators are topological band insulators characterized by a nontrivial bulk gap and pairs of helical edge states protected by time-reversal symmetry or spin conservation[1-4]. Their realization typically relies on sufficiently strong inter-site hopping and spin–orbit coupling, with the chemical potential lying inside a bulk gap[5-15]. In moiré systems, QSH states have generally been predicted and observed at full fillings of spin-degenerate bands, corresponding to an even number of electrons per moiré unit cell[16-24]. By contrast, Mott insulating states arise when electron–electron interactions dominate over kinetic energy. They typically occur at partial band filling, most prominently at half filling of a spin-degenerate band, where an odd number of electrons per moiré unit cell localizes into an insulating state[25-29]. QSH and Mott insulating phases therefore represent opposite limits of the kinetic-to-interaction energy ratio and are conventionally viewed as distinct regimes of quantum matter.

The interplay between QSH topology and strong correlations is one of the central questions in interacting topological matter[1, 2, 30, 31]. In weakly interacting systems, helical edge modes are described by a helical Tomonaga–Luttinger liquid, where correlations suppress edge transport at low temperatures[32-36]. At intermediate interaction strength, QSH order may coexist with symmetry-breaking phases such as charge-density-wave order, ferromagnetism and excitonic insulating phases[12-14, 37]. In the strong-coupling limit, electron interactions can drive transitions into topologically trivial Mott phases; for example, in the Kane–Mele–Hubbard model a transition from a QSH insulator to an antiferromagnetic Mott insulator occurs with increasing Coulomb repulsion[38-41]. Whether a half-filled, correlation-driven insulating state can nevertheless retain robust QSH edge transport remains an outstanding question.

Twisted transition metal dichalcogenides provide a highly tunable platform for addressing this question. Recent experiments on twisted $MoTe_2$ and twisted $WSe_2$ have reported single-, double-, and triple-QSH states at moiré fillings of 2, 4, and 6, respectively[21, 22, 24], as well as signatures of a fractional QSH state with a spin Hall conductivity of 3/2[21]. Unlike conventional QSH systems protected primarily by time-reversal symmetry, these moiré QSH states are protected by spin conservation[21, 22, 42, 43], allowing helical edge modes to remain robust even in the presence of electron–electron interactions and external magnetic fields along the spin quantization axis. Such robustness opens the possibility of realizing strongly correlated insulating phases that coexist with QSH edge transport.

Here, we report the observation of a magnetic-field-stabilized Mott QSH phase hosting a single pair of helical edge states at half filling of the second moiré band in twisted $WSe_2$ (Fig. 1a, b). This phase appears over a broad twist-angle range from 2.29° to 2.89°. At zero out-of-plane magnetic field, combined local and nonlocal transport measurements reveal single and double QSH phases at filling factors $\nu = -2$ and $-4$, respectively. Upon applying an out-of-plane magnetic field of 3 T, a new QSH state emerges at half filling of the second moiré band ($\nu = -3$) and exhibits the same number of edge channels as the $\nu = -2$ single-QSH phase. The helical edge conduction at $\nu = -3$ is further supported by pronounced nonlocal transport and negative in-plane magnetoconductance. Temperature-dependent local transport and Hall measurements reveal a resistance quantization temperature of ~ 5K and Mott gap collapse temperature of ~10 K, corresponding to an energy scale of about 1 meV. These results establish twisted $WSe_2$ as a platform for realizing QSH edge transport in a strongly correlated, half-filled insulating state.

## Characterization of the device at $B_\perp$ = 0T

We first characterize the electrical properties of a quad-gated 2.29° twisted $WSe_2$ device (Fig. 1c) at zero magnetic field. Figure 1d shows the longitudinal sample resistance $R_{xx}$ as a function of moiré filling factor $\nu$ and vertical electric field $E$. The measurements are performed using the local transport geometry illustrated in the inset. The device is biased with a constant current of 10 nA between the source ($I^+$) and drain ($I^-$), and the voltage drop is recorded between $V^+$ and $V^-$. Resistance plateaus/peaks are observed at commensurate fillings $\nu = -1, -2,$ and $-4$. At $\nu = -1$, $R_{xx}$ exceeds 3 MΩ and diverges with decreasing temperatures (Extended Data Fig. 1), indicating a topologically trivial Mott insulating state. In contrast, at $\nu = -2$ and $\nu = -4$, $R_{xx}$ remains comparatively low and is largely insensitive to the vertical electric field over a wide range. These features are consistent with the single QSH and double QSH phases, respectively, in agreement with continuum-model calculations and previous reports in twisted $WSe_2$ near a twist angle of ~3° (Extended Data Fig. 2 and Ref. [22]).

Helical edge conduction at $\nu = -2$ and -4 is further confirmed by the nonlocal transport measurements. Figure 1e shows the nonlocal resistance $R_{nl}$ as a function $\nu$ and $E$. Pronounced $R_{nl}$ is observed near commensurate moiré fillings of $\nu = -1, -2$ and $-4$, while it is negligible for bulk conduction at incommensurate fillings. To exclude geometric or spurious nonlocal contributions, we normalize the nonlocal resistance by the local resistance. Figure 1f shows the nonlocal ratio $R_{nl}/R_{xx}$ as a function of $\nu$ and $E$. For bulk-dominated conduction at non-integer fillings, the ratio is negligible (<3%). For the correlated insulating phase at $\nu = -1$, a sizable nonlocal resistance is observed, but the normalized ratio remains below 1%, consistent with a trivial insulating state. In contrast, the nonlocal ratio is on the order of unity at $\nu = -2$ and $-4$, confirming helical edge-dominated transport for the single ($\nu = -2$) and double QSH ($\nu = -4$) phases in twisted $WSe_2$ (Ref. [22]). Similar nonlocal transport behavior has also been observed in a different measurement geometry (Extended Data Fig. 3).

## Emergence of the QSH phase at $\nu = -3$ and $B_\perp$ = 3T

Next, we examine the electric response of the device under an out-of-plane magnetic field of $B_\perp$ = 3 T. Figure 2a and Figure 2b show the symmetrized longitudinal resistance ( $R_{xx} = \frac{R_{xx}(3\,T) + R_{xx}(-3\,T)}{2}$) and antisymmetrized Hall resistance ($R_{xy} = \frac{R_{xy}(3\,T) - R_{xy}(-3\,T)}{2}$) as a function of $\nu$ and $E$ (see Extended Data Fig. 4 for $R_{xx}$ and $R_{xy}$ at $B_\perp$ = 1, 2 and 4T). The $R_{xx}$ plateaus of the single ($\nu = -2$) and double QSH ($\nu = -4$) phases persist at $B_\perp$ = 3T, accompanied by nearly vanishing $R_{xy}$, consistent with the helical edge-dominated transport in spin-conserved QSH phases. Compared to $B_\perp$ = 0T, a weak resistance plateau emerges near $\nu = -3$, also accompanied by a nearly vanishing Hall response. The odd filling factor of $\nu = -3$ corresponds to the half-filling of the second moiré band and thus a half-filled Mott insulating state. In contrast to the trivial Mott insulator at $\nu = -1$ (half filling of the first moiré band), $R_{xx}$ at $\nu = -3$ is significantly lower than at $\nu = -1$, suggestive of a distinct transport behavior.

To quantitatively understand the $\nu = -3$ state, we study the filling ($\nu$) dependence of $R_{xx}$ (black) and $R_{xy}$ (blue) at $E = 0$ (Fig. 2c). $R_{xx}$ exhibits plateaus/peaks at $\nu = -2, -3$ and -4, with resistance values of about 9.7 kΩ, 9.2 kΩ and 4.5 kΩ, respectively. $R_{xy}$ nearly vanishes at all these fillings. To exclude sample geometry effects and quantitatively compare these three states, we normalize $R_{xx}$ by a reference resistance $R_0$ = 9.7 kΩ ($R_{xx}$ at $\nu = -2$ and $E = 0$). Figure 2d shows the electric field dependence of $R_{xx}/R_0$ at $\nu = -2, -3$ and $-4$. $R_{xx}$ remains approximately constant at ~1.0 $R_0$ for $\nu = -2$ over the studied electric field range, and ~ 0.5 $R_0$ for $\nu = -4$ over $-0.2\ V/nm < E < 0.2\ V/nm$, consistent with single and double QSH phases in the layer-hybridized regime. At $\nu = -3$, $R_{xx}$ also plateaus near ~1.0 $R_0$ over the same electric field range.

The identical $R_{xx}$ plateau at $\nu = -2$ and $-3$, combined with the vanishing $R_{xy}$, suggests that the $\nu = -3$ Mott insulating state is also a QSH state with one pair of helical edge states in the Mott gap. Notably, $R_{xx}$ plateau at $\nu = -3$, nearly identical to that of $\nu = -2$ persists over a wide twist-angle range up to 2.89° (Extended Data Fig. 5).

To confirm helical edge-dominated conduction at $\nu = -3$, we study nonlocal transport under $B_\perp = 3\,T$ in two distinct geometries. As shown in Extended Data Figure 6, pronounced nonlocal resistance $R_{nl}$ appears at $\nu = -3$ for both configurations, providing direct evidence for the edge-dominated transport. The topological protection of the edge states is further supported by the in-plane magneto-conductance. Extended Data Figure 7 shows the in-plane magnetoconductance $\frac{G_{9T} - G_{0T}}{G_{0T}}$ as a function of $\nu$ and $E$. Nearly 30% suppression of electrical conduction is observed near $\nu = -3$, whereas $\frac{G_{9\,T} - G_{0\,T}}{G_{0\,T}}$ is nearly vanishing for bulk-dominated transport at incommensurate fillings. The combined nonlocal and in-plane magnetotransport indicate dominant helical edge conduction at $\nu = -3$ and confirm that it is a QSH insulator with a bulk gap (see Extended Data Fig. 8 for further evidence of the bulk gap from Landau Fan and Hall density analysis near $\nu = -3$).

**Magnetic field dependence of the Mott QSH phase.**

We further investigate the evolution of the $v = -3$ QSH phase under varying out-of-plane magnetic fields. Figure 3a shows $R_{xx}$ as a function of $\nu$ and $B_\perp$ from −7.5 to 7.5 T. The $v = -2$ QSH phase exhibits negligible magnetic field dependence, demonstrating the robust topological protection of helical edge states by spin conservation (Ref. [22]). In contrast, the weaker QSH phases at $v = -3$ and $-4$ emerge only for $B_\perp > 2\,T$. Figure 3b shows the detailed magnetic field dependence of $R_{xx}$ at $v = -2, -3,$ and $-4$. With increasing $B_\perp$, $R_{xx}$ at $v = -3$ and $v = -4$ initially increases and then saturates at approximately 10kΩ and 4.5kΩ, respectively, for $B_\perp > 2\,T$.

To quantitatively compare the different QSH states, we normalize $R_{xx}$ by $R_{v=-2}$. Figure 3c shows $R_{xx}/R_{v=-2}$ as a function of $B_\perp$ at $v = -2, -3$ and $-4$. The ratio $R_{xx}/R_{v=-2}$ remains approximately constant at unity for $v = -2$ throughout the studied magnetic-field range. By contrast, at $v = -3$ and $-4$, $R_{xx}/R_{v=-2}$ is below 10% at $B_\perp = 0$ T, and saturates at approximately 95% and 50%, respectively, for $B_\perp > 2.5\,T$. These observations are consistent with the opening of a Mott gap accompanied by a single pair of helical edge states at $v = -3$ (see Extended Data Fig. 9 for the magnetic field dependence of $R_{nl}$ at $v = -3$). Similar magnetic field dependences of $R_{xx}$ are also observed in devices with larger twist angles (Extended Data Fig. 5).

**Temperature dependence of the Mott QSH phase**

Finally, we investigate the temperature dependence of the $v = -3$ QSH state. Figure 4a shows $R_{xx}$ as a function of $\nu$ at selected temperatures from 1.4 to 15 K. $R_{xx}$ is symmetrized with respect to $B_\perp = \pm 3\,T$. The QSH ($v = -2$) and double QSH ($v = -4$) states exhibit robust resistance peaks up to 15 K, consistent with the temperature-insensitive behavior previously reported in 2.9° twisted $WSe_2$ (Ref. [22]). In contrast, the $v = -3$ QSH state survives only up to $T \sim 10\,K$.

Figure 4b shows the temperature dependence of $R_{xx}/R_0$ at commensurate moiré fillings of $v = -2, -3,$ and $-4$. The $v = -2$ and $v = -4$ states exhibit weak temperature dependence over the entire range from 1.4 to 15 K. By contrast, at $v = -3$, $R_{xx}$ remains plateaued near $0.93R_0$ only for $T < 5\,K$ and decreases rapidly with increasing temperature, reaching approximately 10% of $R_0$ at 15 K. This pronounced temperature dependence suggests a thermally suppressed bulk gap, consistent with a correlated Mott insulating state at half band filling of the moiré band (Ref. [24]).

To further understand the temperature-driven transition at $v = -3$, we investigate the temperature dependence of the Hall response. Figure 5a shows $R_{xy}$ as a function of $\nu$ measured from 1.4 to 15 K. $R_{xy}$ is anti-symmetrized with respect to $B_\perp = \pm 3\,T$. At low temperatures ($T < 10$ K), $R_{xy}$ nearly vanishes at the commensurate filling $v = -3$. $R_{xy}$ is negative on the electron-doped side of the $v = -3$ state and positive on the hole-doped side, consistent with a fully developed Mott gap at $v = -3$. As the temperature increases, $R_{xy}$ on the electron-doped side changes sign.

Figure 5b shows the temperature dependence of $R_{xy}$ at fixed fillings of $v = -2.89$ and $-3.18$, corresponding to electron and hole dopings of 0.11 and -0.18, respectively, relative to $v = -3$. The sign of $R_{xy}$ at $v = -3.18$ remains unchanged throughout the measured temperature range. In contrast, at $v = -2.89$, $R_{xy}$ changes sign from negative to positive at $T^* \approx 10$ K, indicating the disappearance of the electron-like Fermi-surface and collapse of the Mott gap. Figure 5c summarizes $T^*$ as a function of out-of-plane magnetic fields $B_\perp$. $T^*$ remains nearly constant at approximately 10 K for $B_\perp \geqslant 3$ T, indicating that the Mott gap is primarily interaction driven.

## Discussion

We present a qualitative understanding of the magnetic-field-stabilized Mott quantum spin Hall (QSH) phase. Owing to spin conservation (with an out-of-plane spin quantization axis), the helical edge states are largely insensitive to out-of-plane magnetic fields, and the primary effect of $B_\perp$ is on the bulk carriers. As illustrated in Figure 1a, the spin Chern numbers of the first three moiré valence bands in twisted $WSe_2$ are 1, 1, and −2, respectively ($C_s$ = -1, -1, and 2 for the opposite valley, see Extended Data Fig. 2 for details of the band structure). Consequently, the first pair of helical edge states connects the first and third moiré valence bands, while the second pair connects the second and third moiré valence bands. The system therefore realizes a QSH insulator with two helical edge channels when the Fermi energy lies between the first and second moiré valence bands, and a double QSH insulator with four helical edge channels at $v = -4$.

In the weakly interacting regime, at $v = -3$ the second moiré valence band is half filled, and bulk conduction dominates transport, resulting in a metallic state. This is consistent with the absence of a $v = -3$ QSH state at $B_\perp = 0$ T. A moderate out-of-plane magnetic field suppresses bulk conduction through cyclotron motion effects, thereby enhancing the relative importance of electron–electron Coulomb interactions and driving the formation of a correlation-induced bulk gap at $v = -3$. Once this interaction-driven bulk gap opens, the same pair of helical edge states connecting the first and third moiré valence bands dominates electrical transport, giving rise to a Mott quantum spin Hall state.

## Conclusion

In conclusion, we observe a magnetic-field-stabilized Mott quantum spin Hall (QSH) phase at half filling of the second moiré band in a 2.29° twisted $WSe_2$ device. Transport measurements reveal a robust resistance plateau at $v = -3$ emerging above an out-of-plane magnetic field of 3 T and exhibiting the same quantized value as the QSH state at $v = -2$. The helical edge conduction is further confirmed by nonlocal transport and magnetoconductance measurements. Our temperature-dependent measurements indicate a Mott-gap collapse temperature of approximately 10 K. The observed QSH behavior is consistent with the band structure of twisted $WSe_2$, where the emergence of the Mott insulating state at $v = -3$ exposes the underlying helical edge conduction. Our observations demonstrate the persistence of topological edge conduction within a Mott insulating background and establish twisted Transition Metal Dichalcogenides as a promising platform for correlated topological phases.

## References

1. Hasan, M.Z. & Kane, C.L. Colloquium: Topological insulators. *Rev Mod Phys* **82**, 3045–3067 (2010).
2. Qi, X.L. & Zhang, S.C. Topological insulators and superconductors. *Rev Mod Phys* **83** (2011).
3. Chiu, C.K., Teo, J.C.Y., Schnyder, A.P. & Ryu, S. Classification of topological quantum matter with symmetries. *Rev Mod Phys* **88** (2016).
4. Tang, J. et al. Quantum Spin Hall Effects in Van der Waals Materials. *Adv Quantum Technol* **8** (2025).
5. Kane, C.L. & Mele, E.J. Quantum spin Hall effect in graphene. *Phys Rev Lett* **95** (2005).
6. Kane, C.L. & Mele, E.J. topological order and the quantum spin Hall effect -: art. no. 146802. *Phys Rev Lett* **95** (2005).
7. Bernevig, B.A. & Zhang, S.C. Quantum spin hall effect. *Phys Rev Lett* **96** (2006).
8. Konig, M. et al. Quantum spin hall insulator state in HgTe quantum wells. *Science* **318**, 766–770 (2007).
9. Knez, I. & Du, R.R. Quantum spin Hall effect in inverted InAs/GaSb quantum wells. *Front Phys-Beijing* **7**, 200–207 (2012).
10. Fei, Z.Y. et al. Edge conduction in monolayer WTe. *Nat Phys* **13**, 677 (2017).
11. Wu, S.F. et al. Observation of the quantum spin Hall effect up to 100 kelvin in a monolayer crystal. *Science* **359**, 76–79 (2018).
12. Zhao, W.J. et al. Magnetic proximity and nonreciprocal current switching in a monolayer $WTe_2$ helical edge. *Nat Mater* **19**, 503(2020).
13. Tang, J. et al. Dual quantum spin Hall insulator by density-tuned correlations in $TaIrTe_4$. *Nature* **628**, 515 (2024).
14. Tang, J. et al. Bistable superlattice switching in a quantum spin Hall insulator. *Nature* **652** (2026).
15. Liu, X.Y. et al. Prediction of the Dual Quantum Spin Hall Insulator in the $NbIrTe_4$ Monolayer. *Chinese Phys Lett* **42** (2025).
16. Wu, F.C., Lovorn, T., Tutuc, E., Martin, I. & MacDonald, A.H. Topological Insulators in Twisted Transition Metal Dichalcogenide Homobilayers. *Phys Rev Lett* **122** (2019).
17. Devakul, T., Crépel, V., Zhang, Y. & Fu, L. Magic in twisted transition metal dichalcogenide bilayers. *Nat Commun* **12** (2021).

18. Xu, C., Li, J.X., Xu, Y., Bi, Z. & Zhang, Y. Maximally localized Wannier functions, interaction models, and fractional quantum anomalous Hall effect in twisted bilayer MoTe2. *P Natl Acad Sci USA* **121** (2024).
19. Li, T.X. et al. Quantum anomalous Hall effect from intertwined moire bands. *Nature* **600**, 641–+ (2021).
20. Foutty, B.A. et al. Mapping twist-tuned multiband topology in bilayer WSe. *Science* **384**, 343–347 (2024).
21. Kang, K.F. et al. Evidence of the fractional quantum spin Hall effect in moiré MoTe. *Nature* **628** (2024).
22. Kang, K.F. et al. Double Quantum Spin Hall Phase in Moire WSe. *Nano Lett* **24**, 14901–14907 (2024).
23. Zhao, W.J. et al. Realization of the Haldane Chern insulator in a moiré lattice. *Nat Phys* **20** (2024).
24. Xu, F. et al. Interplay between topology and correlations in the second moiré band of twisted bilayer MoTe. *Nat Phys* **21** (2025).
25. Imada, M., Fujimori, A. & Tokura, Y. Metal-insulator transitions. *Rev Mod Phys* **70**, 1039–1263 (1998).
26. Cao, Y. et al. Correlated insulator behaviour at half-filling in magic-angle graphene superlattices. *Nature* **556**, 80 (2018).
27. Chen, G.R. et al. Evidence of a gate-tunable Mott insulator in a trilayer graphene moire superlattice. *Nat Phys* **15**, 237–241 (2019).
28. Tang, Y.H. et al. Simulation of Hubbard model physics in $WSe_2$/$WS_2$ moire superlattices. *Nature* **579**, 353 (2020).
29. Li, T.X. et al. Continuous Mott transition in semiconductor moire superlattices. *Nature* **597**, 350 (2021).
30. Maciejko, J., Hughes, T.L. & Zhang, S.C. The Quantum Spin Hall Effect. *Annu Rev Conden Ma P* **2**, 31–53 (2011).
31. Stern, A. Fractional Topological Insulators: A Pedagogical Review. *Annual Review of Condensed Matter Physics, Vol 7* **7**, 349–368 (2016).
32. Wu, C.J., Bernevig, B.A. & Zhang, S.C. Helical liquid and the edge of quantum spin Hall systems. *Phys Rev Lett* **96** (2006).
33. Xu, C.K. & Moore, J.E. Stability of the quantum spin Hall effect: Effects of interactions, disorder, and $Z_2$ topology. *Phys Rev B* **73** (2006).
34. Li, T.X., Wang, P.J., Sullivan, G., Lin, X. & Du, R.R. Low-temperature conductivity of weakly interacting quantum spin Hall edges in strained- layer InAs/GaInSb. *Phys Rev B* **96** (2017).
35. Stühler, R. et al. Tomonaga-Luttinger liquid in the edge channels of a quantum spin Hall insulator. *Nat Phys* **16**, 47 (2020).
36. Jia, J.X. et al. Tuning the many-body interactions in a helical Luttinger liquid. *Nat Commun* **13** (2022).
37. Jia, Y.Y. et al. Evidence for a monolayer excitonic insulator. *Nat Phys* **18**, 87 (2022).
38. Hohenadler, M. et al. Quantum phase transitions in the Kane-Mele-Hubbard model. *Phys Rev B* **85** (2012).
39. Vaezi, A., Mashkoori, M. & Hosseini, M. Phase diagram of the strongly correlated Kane-Mele-Hubbard model. *Phys Rev B* **85** (2012).

40. Hung, H.H., Wang, L., Gu, Z.C. & Fiete, G.A. Topological phase transition in a generalized Kane-Mele-Hubbard model: A combined quantum Monte Carlo and Green's function study. *Phys Rev B* **87** (2013).
41. Meng, Z.Y., Hung, H.H. & Lang, T.C. The Characterization of Topological Properties in Quantum Monte Carlo Simulations of the Kane-Mele-Hubbard Model. *Mod Phys Lett B* **28** (2014).
42. Liu, L., Liu, Y.T., Li, J.Y., Wu, H. & Liu, Q.H. Quantum spin Hall effect protected by spin U(1) quasisymmetry. *Phys Rev B* **110** (2024).
43. Liu, L., Liu, Y.T., Li, J.Y., Wu, H. & Liu, Q.H. Orbital doublet driven even-spin Chern insulators. *Phys Rev B* **110** (2024).
44. Wang, L. et al. One-Dimensional Electrical Contact to a Two-Dimensional Material. *Science* **342**, 614–617 (2013).

## Methods

### Device fabrication

The twisted $WSe_2$ quad-gate devices were fabricated using the tear-and-stack and layer-by-layer transfer method[44]. First, $WSe_2$, hexagonal boron nitride (hBN) and few-layer graphite flakes were mechanically exfoliated using adhesive tapes and identified by optical contrast. Monolayer $WSe_2$ flakes were then cut into two pieces using an AFM tip to obtain two monolayer flakes with a well-defined relative twist angle. The heterostructures were assembled using a dry transfer stamp consisting of a thin polycarbonate (PC) film on polydimethylsiloxane (PDMS). Using this stamp, we sequentially picked up a thin hBN flake, one monolayer $WSe_2$ piece, a second monolayer $WSe_2$ piece at a controlled small twist angle relative to the first, another hBN flake (narrower than the $WSe_2$ flake), and finally a few-layer graphite flake (2–4 μm wide). The resulting hBN/$WSe_2$/$WSe_2$/hBN/graphite stack was then released onto a Si/$SiO_2$ substrate pre-patterned with platinum Hall bar electrodes. Rectangular local contact gates were defined using standard electron-beam lithography and subsequently coated with palladium. Finally, a wider few-layer graphite flake and an additional hBN flake were sequentially picked up and transferred onto the contact gates to complete the device architecture. The final twist angle of the device was determined by the Landau level degeneracies following Ref. [20].

### Electrical measurements

The electrical measurements were performed in a closed cycle $^4$He cryostat (Physike VTI) equipped with a 9T superconducting magnet. Low-frequency (<23 Hz) lock-in techniques were used to measure the sample resistance under a small bias current (<10 nA) to minimize sample heating. Both the source-drain current and the voltage drop at the probe electrode pairs were simultaneously recorded. For the local transport measurements, the device was biased symmetrically along the major axis of the Hall bar to ensure a uniform current flow (see inset of Fig. 1d). For the nonlocal measurements, the current was injected perpendicular to the major axis (inset of Fig. 1e), forming an effective H-bar geometry. The validity of this nonlocal configuration was verified by normalizing the nonlocal resistance $R_{nl}$ to the longitudinal resistance $R_{xx}$. We found that $R_{nl}$ is suppressed to below 3% of $R_{xx}$ in the bulk conducting regime at incommensurate moiré fillings, and to < 1% in the trivial Mott insulating regime (Fig. 1f).

**Continuum Model**

The moiré band structure of small-angle-twisted $WSe_2$ bilayers was calculated using a continuum model following Ref. [16] . The Hamiltonian for the K valley states of $WSe_2$ is given by

$$H_K = \begin{pmatrix} \frac{\hbar^2 k^2}{2m^*} + \Delta_t(\boldsymbol{r}) + \frac{V_z}{2} & \Delta_T(\boldsymbol{r}) \\ \Delta_T^\dagger(\boldsymbol{r}) & \frac{\hbar^2 k^2}{2m^*} + \Delta_b(\boldsymbol{r}) - \frac{V_z}{2} \end{pmatrix}$$

Here $\frac{\hbar^2 k^2}{2m^*}$ is the kinetic energy with $k$ and $m*$ denoting the crystal momentum and effective electron mass; $\Delta_{t,b}(\boldsymbol{r})$ represent intralayer moiré potential for the top and bottom layers, $\Delta_T(\boldsymbol{r})$ represents the interlayer tunneling amplitude; $V_z$ is the interlayer electrical potential difference tunable by a vertical electric field. In the calculation, we used an effective electron mass of $m^* = 0.43m_e$, an intralayer moiré potential with an amplitude of $V$ = -9 meV and phase of -128 °, and an interlayer tunneling amplitude of -18 meV. The Hamiltonian was diagonalized using the plane-wave basis and only < $5^{th}$ shell of the moiré Brillouin zone is retained. The spin/valley Chern number of each moiré band was obtained by integrating the Berry curvatures within the first Brillouin zone. The Hamiltonian for the -K valley states is a time-reversed copy.

**Author contributions:**

K.K. conceived the project. Y.J., Y.M. and A.Z., fabricated the devices with the assistance of N.Z.. K.K., U. L., I.W. and T.K.C. set up the measurement equipment. K.K., Y.J., Y.M. and A.Z. performed the electrical measurements. K.K. analyzed data, calculated the band structure and wrote the manuscript with input from all the authors.

**Acknowledgements**

We acknowledge helpful discussions with Kin Fai Mak, Xi Dai, Wei Zhu and Adrian Hoi Chun PO. This research is mainly supported by the Ministry of Science and Technology (Project No. 2025YFE0201200) awarded to K.K.. N. W. acknowledges financial support from the Research Grants Council (RGC) of Hong Kong (Project No. C6053-23G). N.W. and K.K. also acknowledge support from (Project No. AoE/P-604/25R Awarded to N.W. and K.K.) for novel device fabrications in this research.

## Main Figures

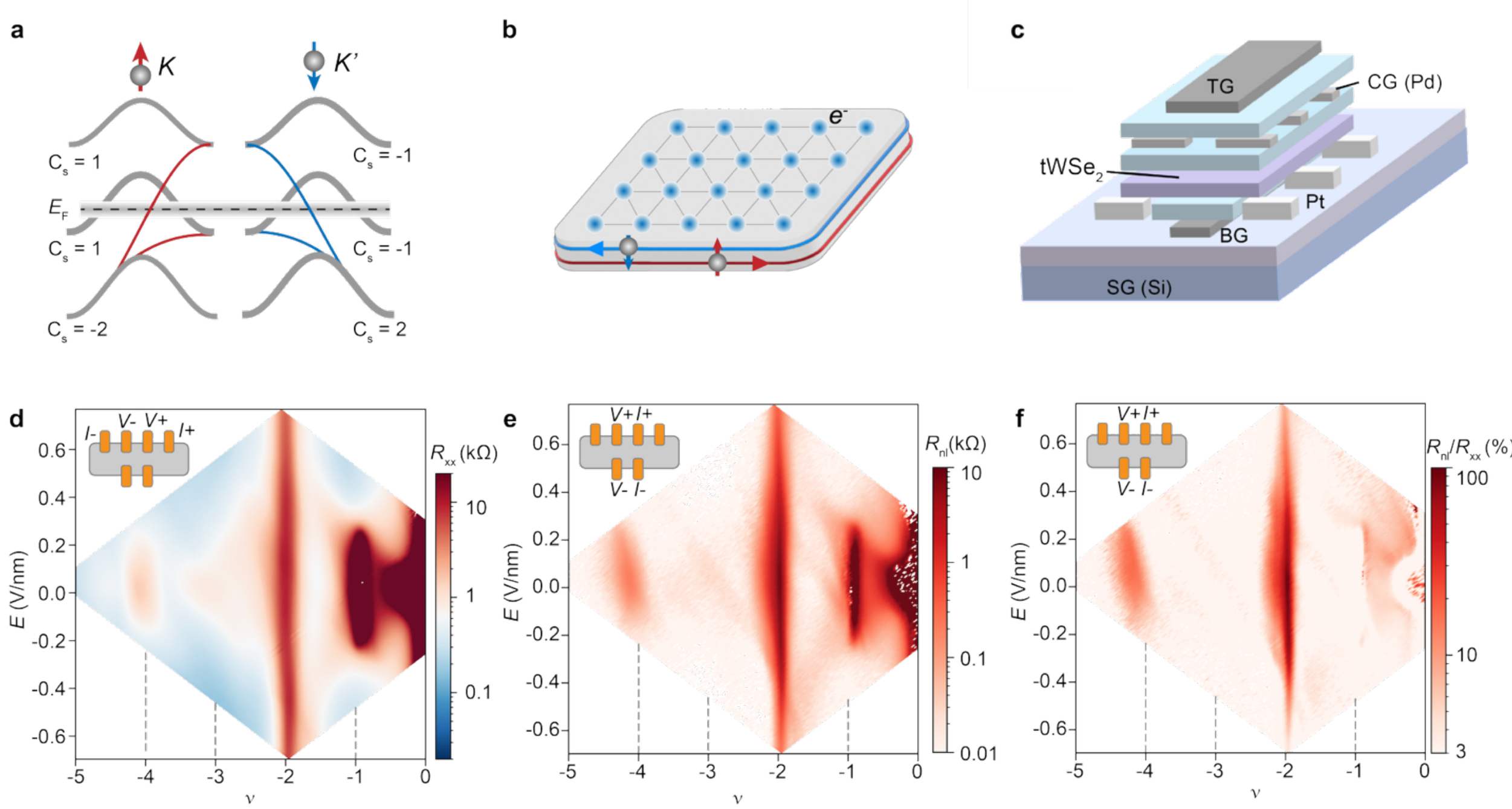

**Figure 1 | Mott quantum spin Hall phase in twisted $WSe_2$. a,** Schematics of the Mott quantum spin Hall (QSH) state. At half filling of the second moiré band (black dashed line denotes the Fermi energy, $E_F$), strong electron correlations open a Mott gap. Two helical edge states (blue and red lines for different spin/valley) emerge in each valley: the first connects the first and third moiré valence bands, while the second connects the second and third moiré valence bands. **b**, Real-space illustration of the Mott QSH phase. In the bulk, carriers (blue dots) are localized on the triangular moiré lattice by strong Coulomb repulsion, forming a Mott insulating state. Spin-momentum–locked helical edge states propagate along the edge of the Mott insulator. **c,** Device architecture of a quad-gated twisted $WSe_2$ device. Graphite top (TG) and bottom (BG) gates independently control carrier density and vertical displacement field. Palladium contact gates (CG) provide strong hole doping to reduce contact resistance at the platinum (Pt) electrodes, while a silicon depletion gate (SG) removes excess carriers outside of the overlap region of TG and BG. **d-f,** Longitudinal resistance $R_{xx}$, nonlocal resistance $R_{nl}$ and the nonlocal ratio $R_{nl}/R_{xx}$ as functions of moiré filling factor $\nu$ and vertical electric field *E,* respectively. Insets show the corresponding measurement configurations, where $I^+$ and $I^-$ denote source and drain contacts, and $V^+$ and $V^-$ denote voltage probes. All measurements are at 1.4 K.

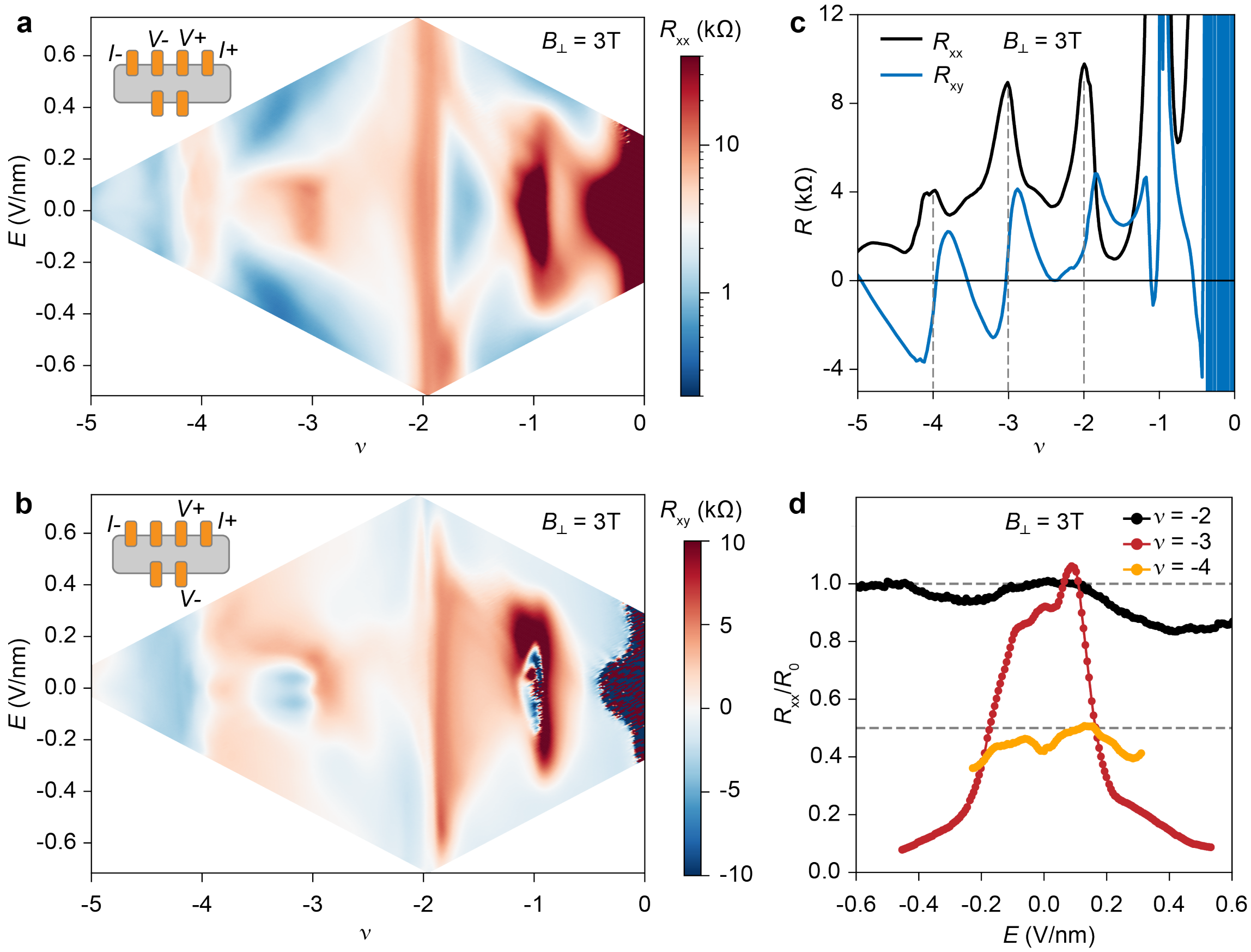

**Figure2| Emergence of the quantum spin Hall phase at $\nu = -3$ and $B_\perp = 3$ T. a, b,** Longitudinal sample resistance $R_{xx}$ (**a**) and Hall resistance $R_{xy}$ (**b**) as a function of $\nu$ and $E$. $R_{xx} = \frac{R_{xx}(3\ T) + R_{xx}(-3\ T)}{2}$ is symmetrized and $R_{xy} = \frac{R_{xy}(3\ T) - R_{xy}(-3\ T)}{2}$ is antisymmetrized at $B_\perp = \pm 3$T respectively. **c,** $R_{xx}$ (black) and $R_{xy}$ (blue) as a function of $\nu$ at $B_\perp$ = 3T. **d,** Normalized longitudinal sample resistance $R_{xx}/R_0$ as a function of the vertical electric field at selected filling factors of $\nu = -2, -3$ and $-4$. Here $R_0 = 9.7$ kΩ is $R_{xx}$ of the device at $\nu = -2$ and $E = 0$. All measurements are at 1.4 K.

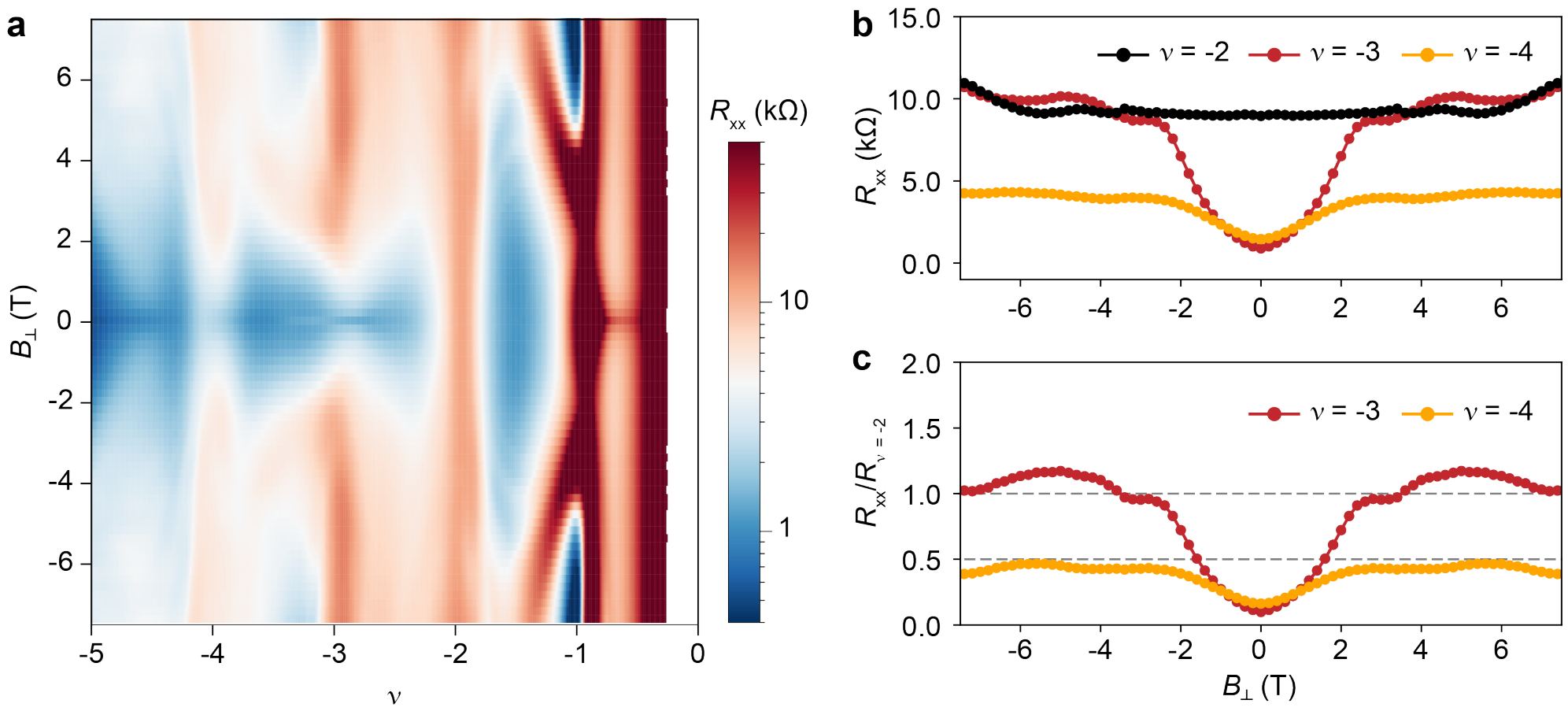

**Figure3| Magnetic field dependence of the quantum spin Hall effect at $\nu = -3$. a,** $R_{xx}$ as a function of the out-of-plane magnetic field $B_{\perp}$ and moiré filling factor $\nu$. **b,** $R_{xx}$ as a function of $B_{\perp}$ at selected moiré filling factor of $\nu = -2, -3$, and $-4$. **c,** Magnetic field dependence of the normalized sample resistance $\frac{R_{xx}}{R_{\nu=-2}}$ for $\nu = -3$ and $-4$, where $R_{\nu=-2}$ is the plateau resistance of the well-established QSH state at $\nu = -2$. The evolution shows that increasing $B_{\perp}$ stabilizes the QSH phase at $\nu = -3$, with its resistance approaching that of the $\nu = -2$ state at sufficiently high fields. All measurements are at 1.4 K.

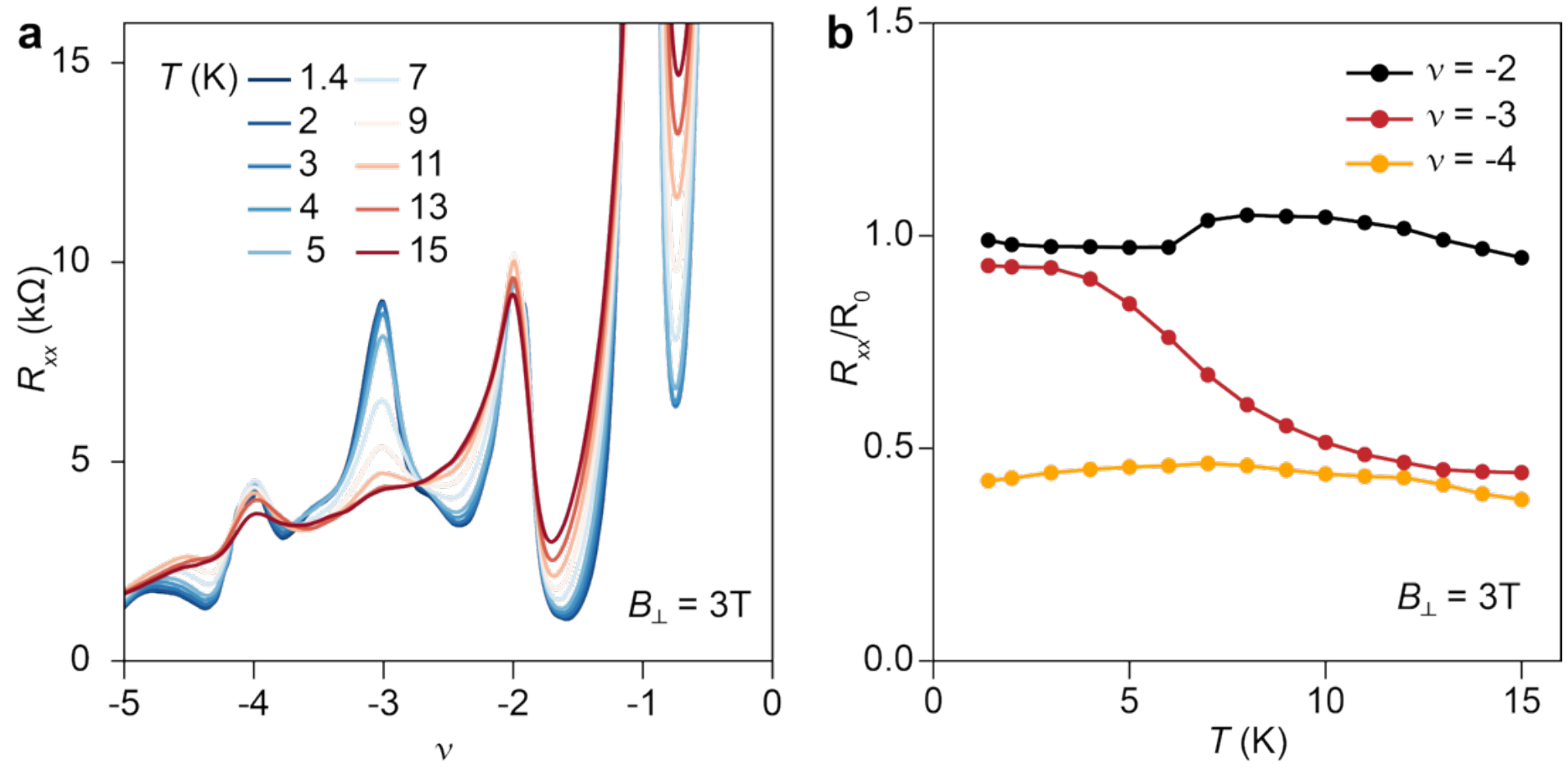

**Figure4| Temperature dependence of the Mott quantum spin Hall phase. a,** Symmetrized longitudinal sample resistance $R_{xx} = \frac{R_{xx}\,(3\,T) - R_{xx}\,(-3\,T)}{2}$ as a function of $\nu$ at selected temperatures from 1.4 to 15 K. **b,** Normalized longitudinal sample resistance $R_{xx}/R_0$ at selected moiré fillings of $\nu = -2, -3$, and $-4$. Here $R_0 = 9.7$ kΩ is the quantized plateau resistance at $\nu = -2$.

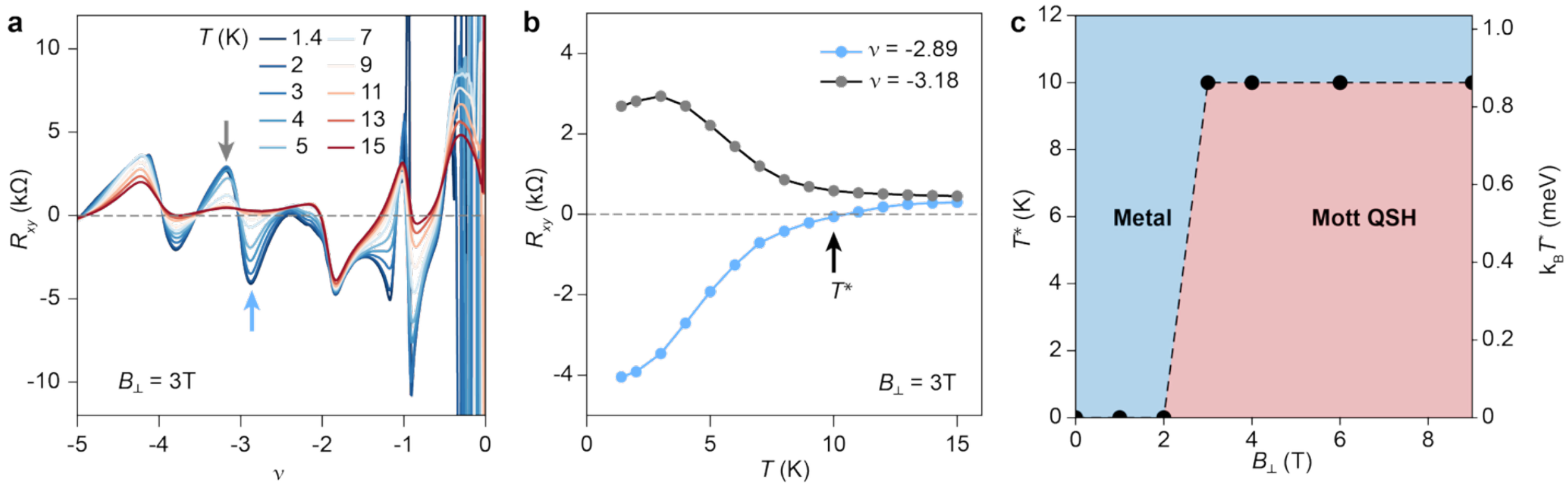

**Figure 5| Phase diagram of the Mott quantum spin Hall phase. a,** Antisymmetrized Hall resistance $R_{xy} = \frac{R_{xy}(3\,T)-R_{xy}(-3\,T)}{2}$ at selected temperatures from 1.4 to 15 K. **b,** Temperature dependence of $R_{xy}$ at selected fillings of $\nu = -2.89$ and $\nu = -3.18$; the corresponding filling selections are indicated by the arrows of matching colors in **c.** $T^*$ denotes the temperature at which $R_{xy}$ changes sign for $\nu = -2.89$, marking disappearance of the electron-like Fermi pocket. **c,** Left axis: $T^*$ (black circles) as a function of $B_\perp$. Right axis: the corresponding energy of $k_B T^*$. The dashed line delineates the phase boundary between the metallic phase (blue) and the Mott quantum spin Hall phase (red).

## Extended Data Figures

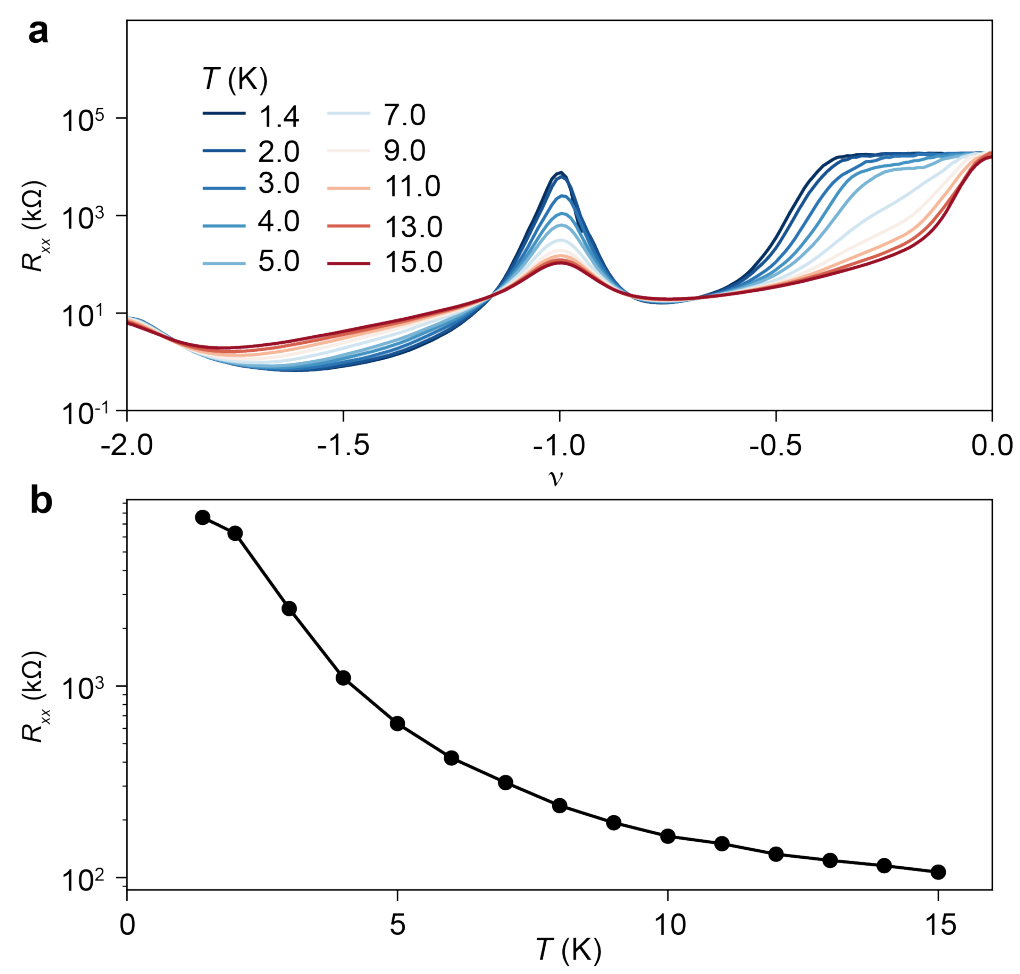

**Extended Data Figure 1| Temperature dependence of the $\nu = -1$ trivial Mott insulating state. a,** $R_{xx}$ as a function of $\nu$ at selected temperatures from 1.4 to 15 K. **b,** Temperature dependence of the $R_{xx}$ at $\nu = -1$. $R_{xx}$ increases exponentially upon cooling, indicating thermally activated transport with a charge gap opening at low temperature. This behavior is consistent with a topologically trivial Mott insulating phase, where conduction is suppressed in the bulk and transport is governed by activation across the interaction-induced gap.

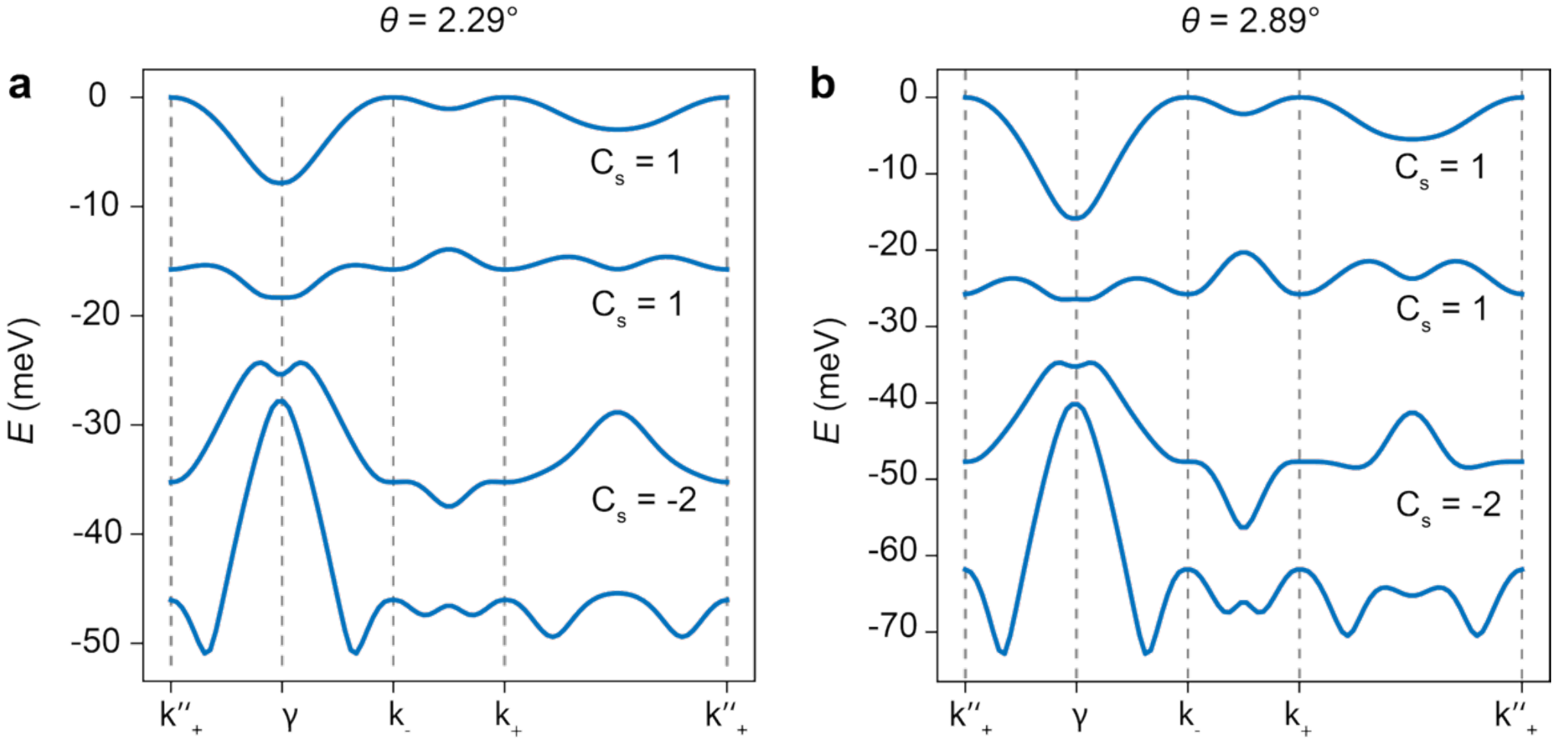

**Extended Data Figure 2| Continuum model band structures of the 2.29° (a) and 2.89° (b) twisted $WSe_2$ in the K valley.** The band structures for the two twist angles are qualitatively similar, indicating that the low-energy moiré physics is robust against small variations in the twist angle. In both cases, the top three valence moiré bands carry nonzero spin Chern numbers, given by $C_s$ =

1,1, and −2, respectively. This topological structure is preserved across the two twist angles, highlighting the stability of the underlying moiré band topology in twisted $WSe_2$.

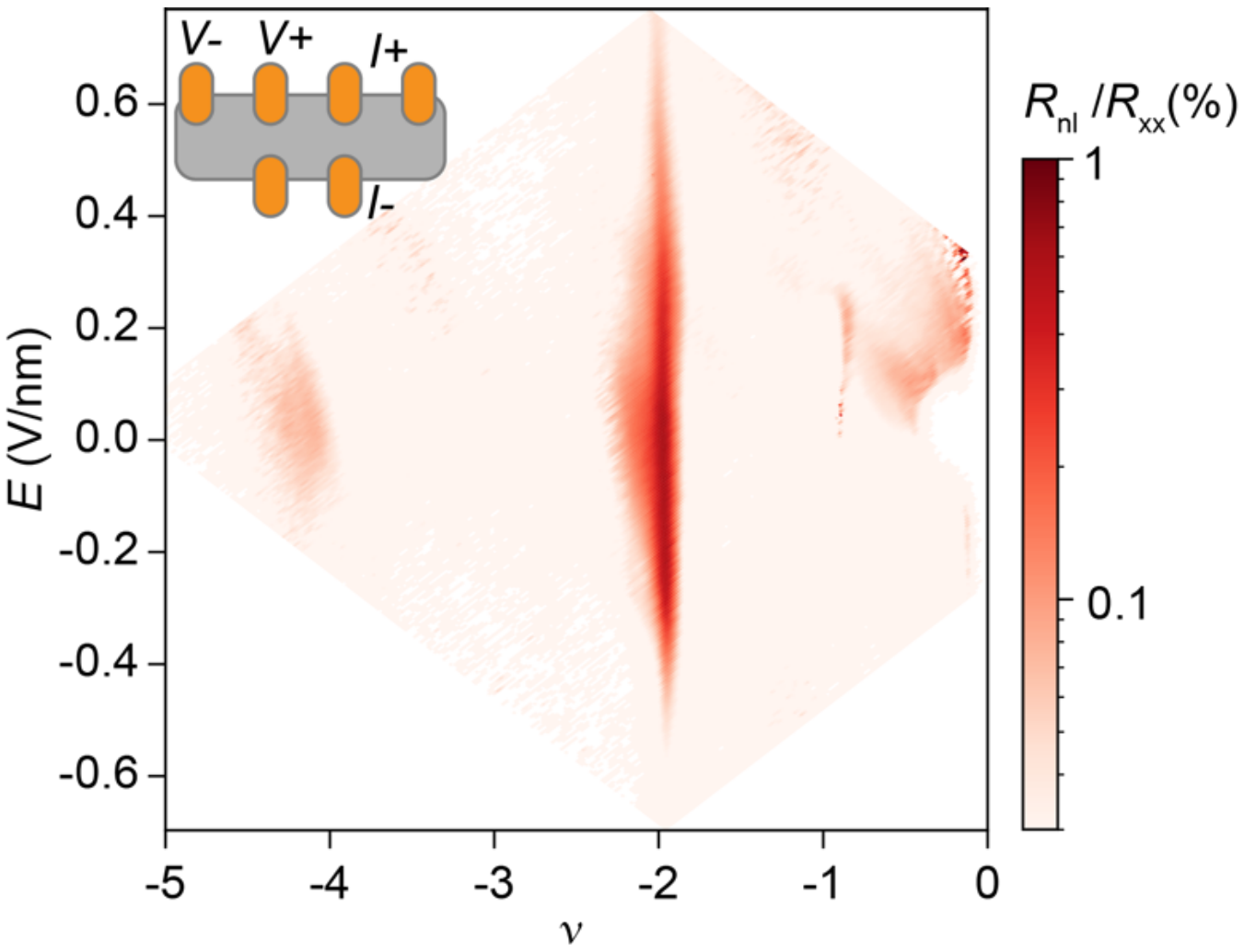

**Extended Data Figure 3|** The nonlocal ratio $R_{nl}/R_{xx}$ as functions of moiré filling factor $\nu$ and vertical electric field $E$, measured at 1.4 K. The inset shows the corresponding measurement configurations, where $I^+$ and $I^-$ denote source and drain contacts, and $V^+$ and $V^-$ denote voltage probes.

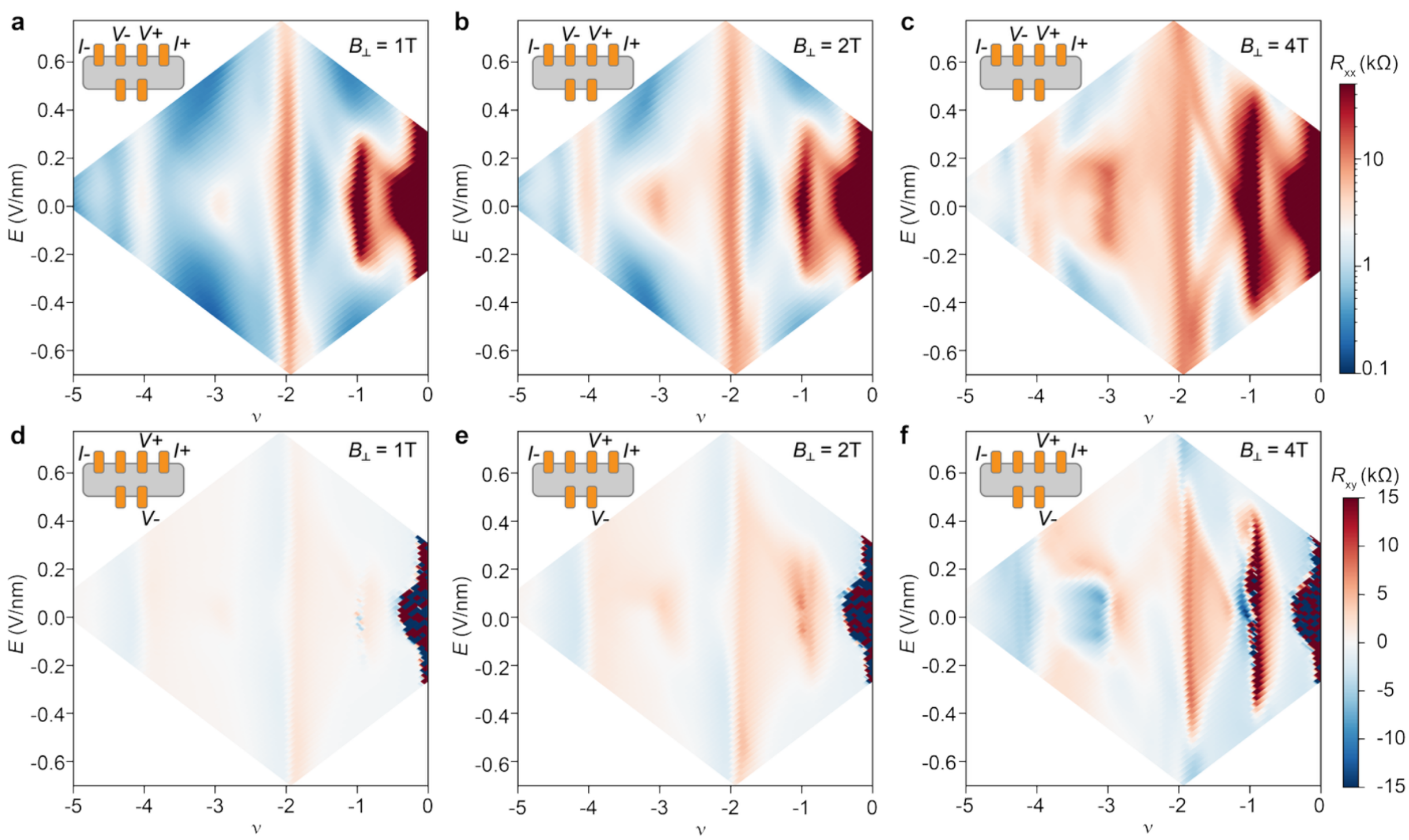

**Extended Data Figure 4| Evolution of $R_{xx}$ (top panel) and $R_{xy}$ (bottom panel) under varying $B_{\perp}$. a-c,** Magnetic field dependence of $R_{xx}$ at selected $B_{\perp}$ of 1, 2 and 4 T, $R_{xx}$ is symmetrized with respect to magnetic field. **d-f,** Magnetic field dependence of $R_{xy}$ at the same selected fields; $R_{xy}$ is anti-symmetrized to remove longitudinal mixing contributions. As $B_{\perp}$ increases from 1 T to 4 T, an additional $R_{xx}$ peak develops at $\nu = -3$, while $R_{xy}$ vanishes. At $B_{\perp} = 4$ T, $R_{xx}$ becomes nearly identical between $\nu = -2$ and $\nu = -3$.

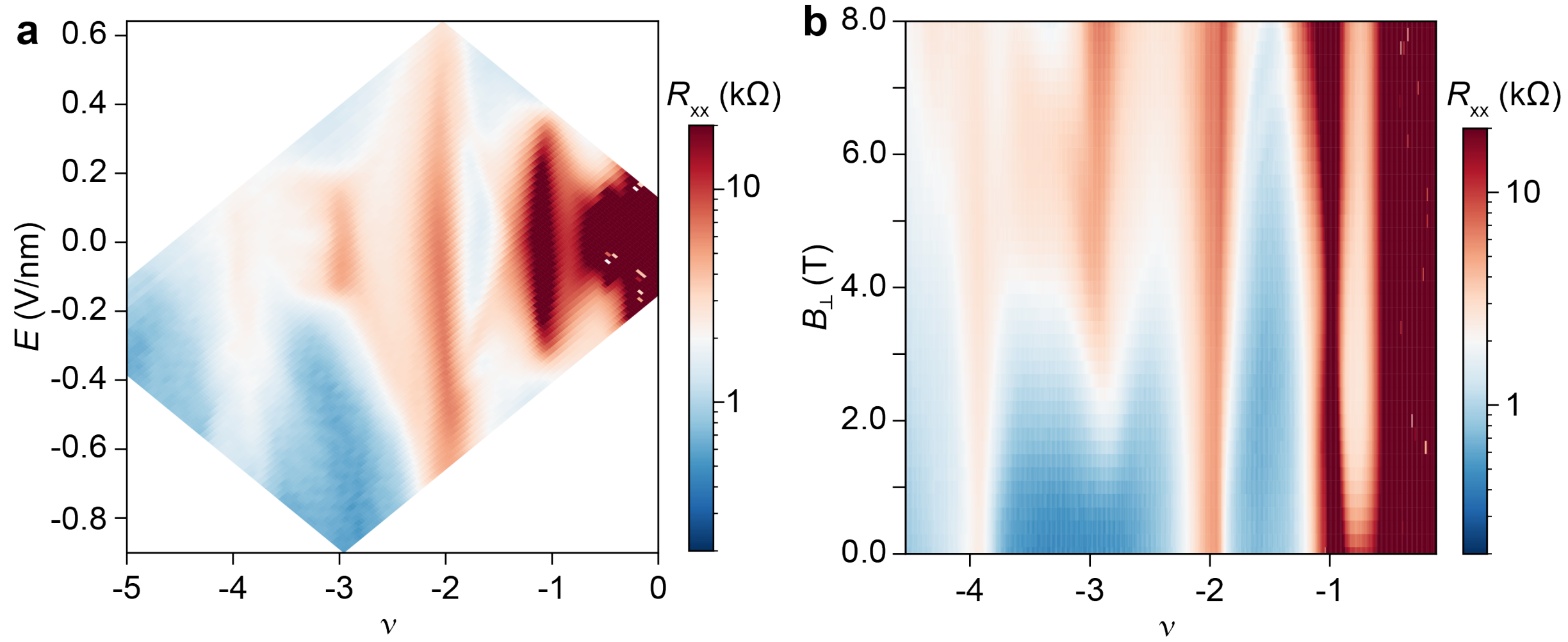

**Extended Data Figure 5| Transport data of an additional twisted $WSe_2$ device with a twisting angle of 2.89°. a,** $R_{xx}$ as a function of moiré filling factor $\nu$ and vertical electric field $E$. $R_{xx}$ is symmetrized and $R_{xy}$ is antisymmetrized using data taken at $B_{\perp} = \pm 5$ T. $R_{xx}$ exhibits plateaus at both $\nu = -2$ and $\nu = -3$, with nearly identical value of ~6 kΩ; the reduced plateau resistance likely arises from imperfect current distribution and residual bulk conduction. **b,** $R_{xx}$ as a function of $\nu$ and $B_{\perp}$ at $E = 0$ V/nm. At $B_{\perp} = 0$ T, only the single ($\nu = -2$) and double ($\nu = -4$) QSH phases are observed. Upon increasing the magnetic field to $B_{\perp} \approx 2$ T, a new insulating state emerges at $\nu = -3$. Above $B_{\perp} = 4$ T, $R_{xx}$ at $\nu = -2$ and $-3$ becomes nearly identical and remains unchanged upon further increase of $B_{\perp}$. These observations are consistent with the magnetic-field-stabilized QSH phase at $\nu = -3$, similar to that observed in the 2.29° twisted $WSe_2$ device. The slightly larger saturation magnetic field is likely due to increased bandwidth and carrier densities at $\nu = -3$ in the larger twist-angle device.

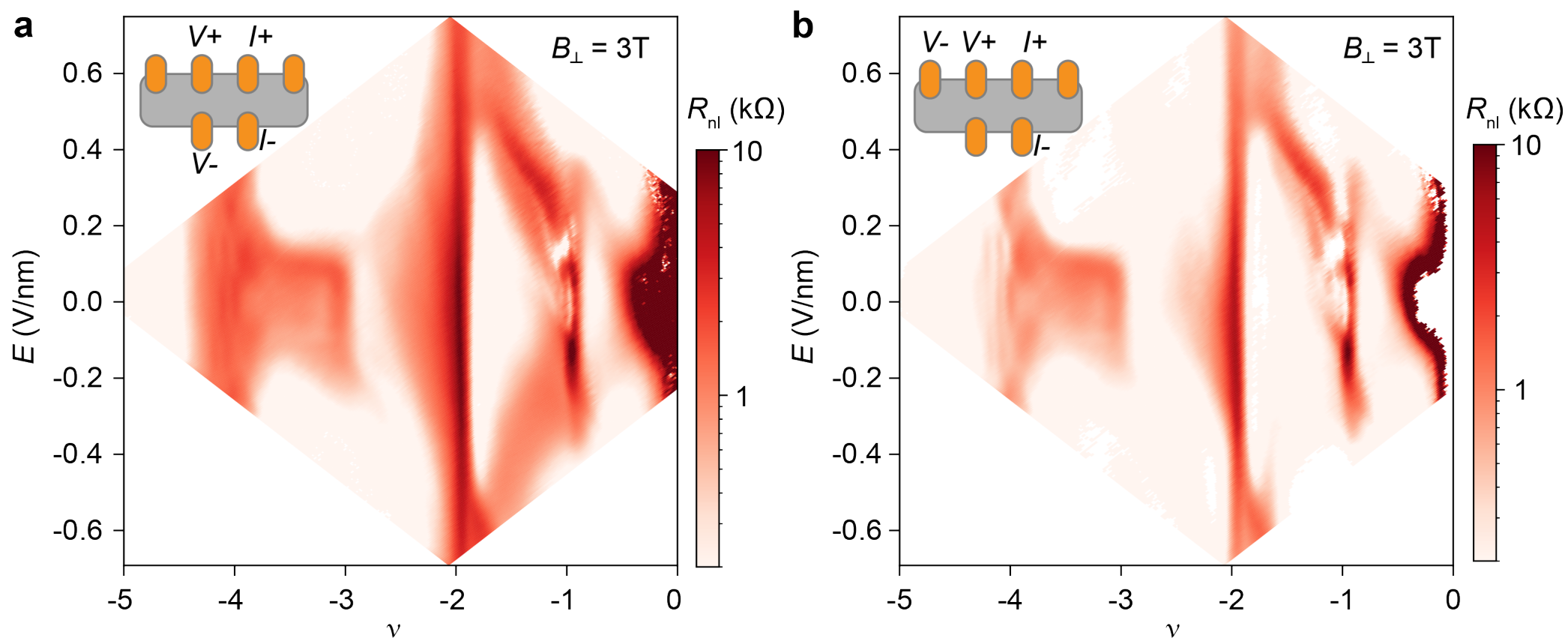

**Extended Data Figure 6| $R_{nl}$ as a function of $v$ and E in two nonlocal measurement configurations (The inset shows the electrical connections).** For both configurations, pronounced nonlocal signals are observed at commensurate fillings of $v = -2, -3$ and $-4$, providing direct evidence of edge-dominated conduction at these fillings. The reduced $R_{nl}$ at $v = -3$ compared to $v = -2$ is likely due to the reduced bulk gap size: the $v = -3$ QSH gap is ~ 1 meV from the temperature dependence, while $v = -2$ QSH gap is ~ 6 meV (Ref. [22]).

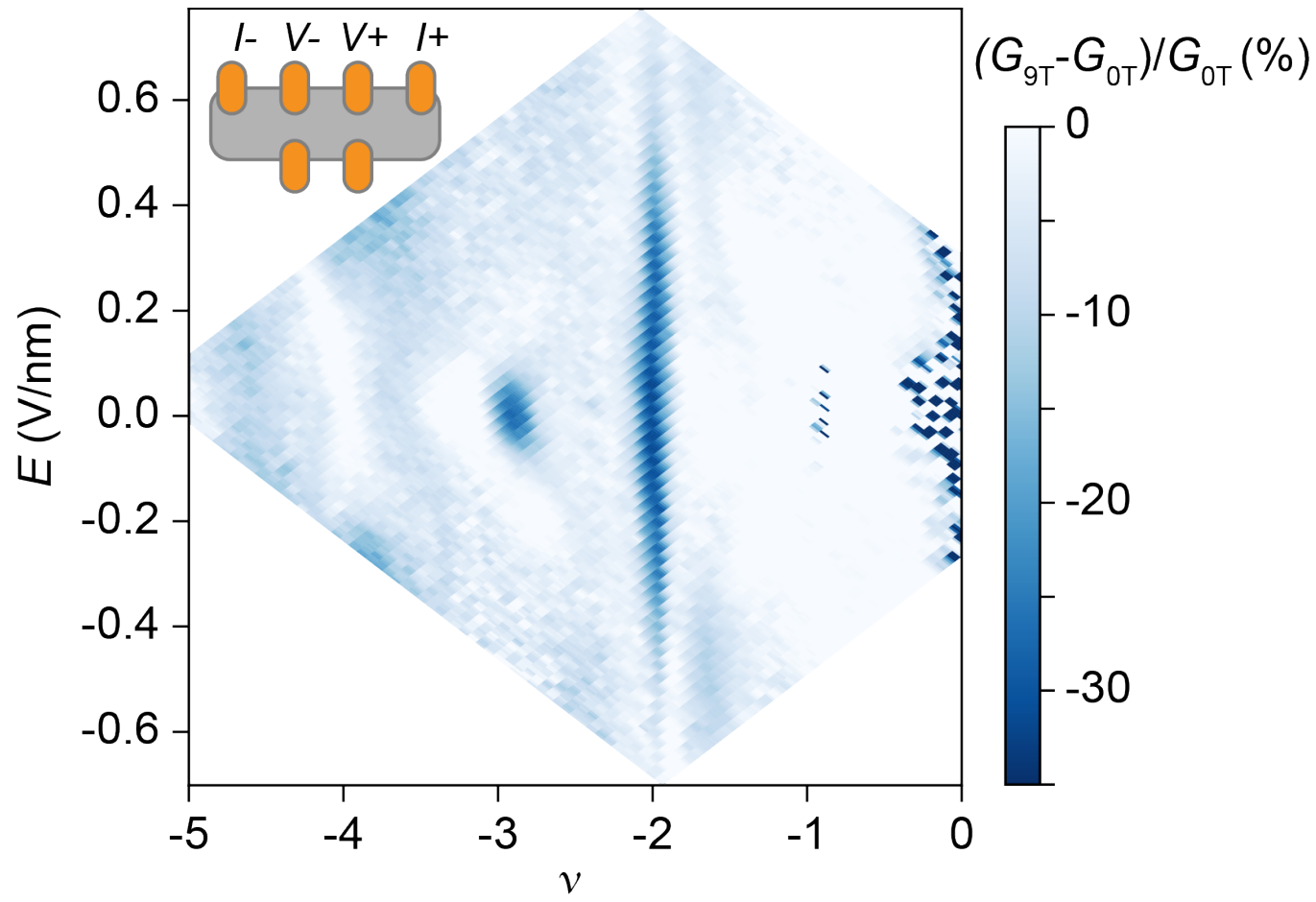

**Extended Data Figure 7|In-plane magneto-conductance $\frac{G_{9T} - G_{0T}}{G_{0T}}$ as a function of $v$ and $E$.** Device conductance shows clear reductions of nearly 30% at $v = -2$ and $-3$, providing direct evidence for helical-edge-dominated conduction that is protected by Ising spin conservation (Ref. [21]).

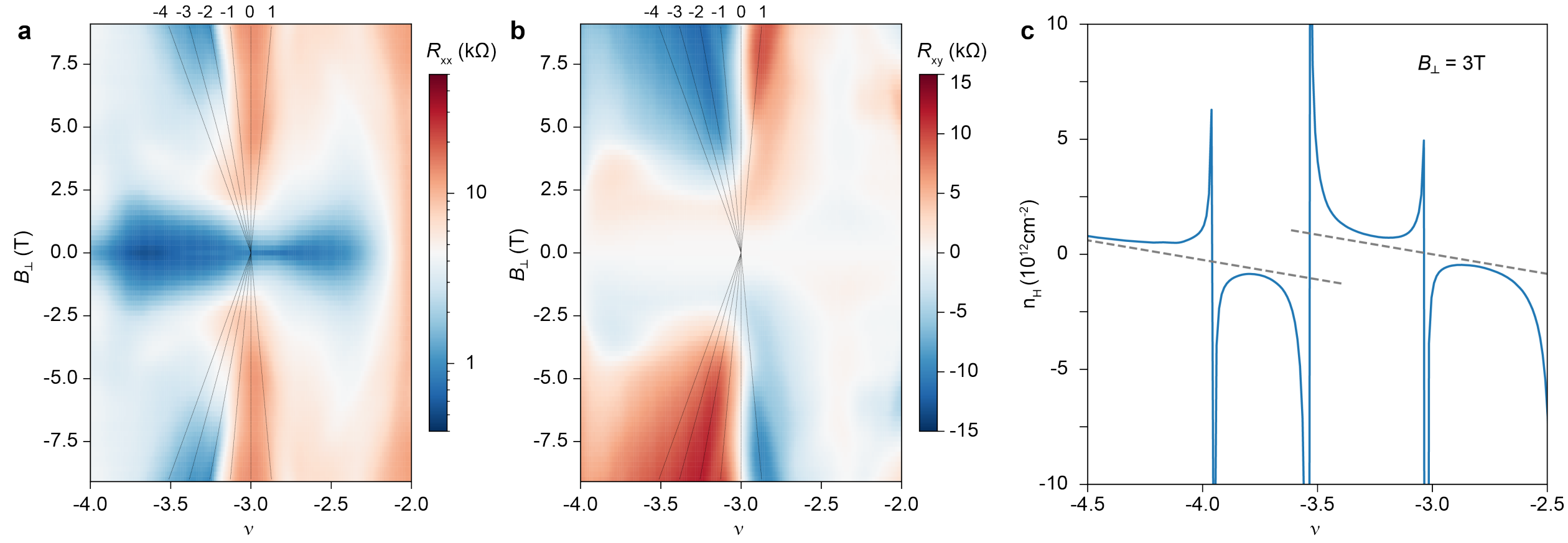

**Extended Data Figure 8| Landau Fan and Hall density near $v = -3$ phase at $T = 1.4$K.** a, b $R_{xx}$ and $R_{xy}$ as functions of $v$ and $B_\perp$. For $B_\perp < 2T$, $R_{xy}$ retains the same sign as $v$ crosses $v = -3$, consistent with the metallic $v = -3$ phase. For $B_\perp > 2.5T$, the sign of $R_{xy}$ reverses as $\nu$ crosses $v = -3$, indicating a transition between hole-like and electron-like carriers separated by a bulk gap. Clear dips in $R_{xx}$ and peaks in $R_{xy}$ emerge with Landau level indices of $v_{LL}$ = -4, -3, -2, -1 and 1. Their corresponding moiré filling factor $\nu$ varies linearly with $B_\perp$ following the Streda formula (black dashed lines) . Both the hole-like Landau levels ($v_{LL}$ = -4, -3, -2, -1) and electron-like Landau levels ($v_{LL}$ = 1) converge toward $v = -3$ as $B_\perp \rightarrow 0$T. These observations indicate the opening of a bulk gap at $v = -3$. **c**, Hall density $n_H$ as a function of $\nu$ measured at $B_\perp$ = 3 T. In contrast to $v = -3.5$, where a van Hove singularity produces an abrupt Hall density jump of approximately 2$n_{moire}$ ($n_{moire}$ is the moiré density), $n_H$ linearly sweeps through $v = -3$ with no such jumps. This behavior is consistent with a QSH phase featuring a bulk gap and helical edge states, rather than a van Hove singularity.

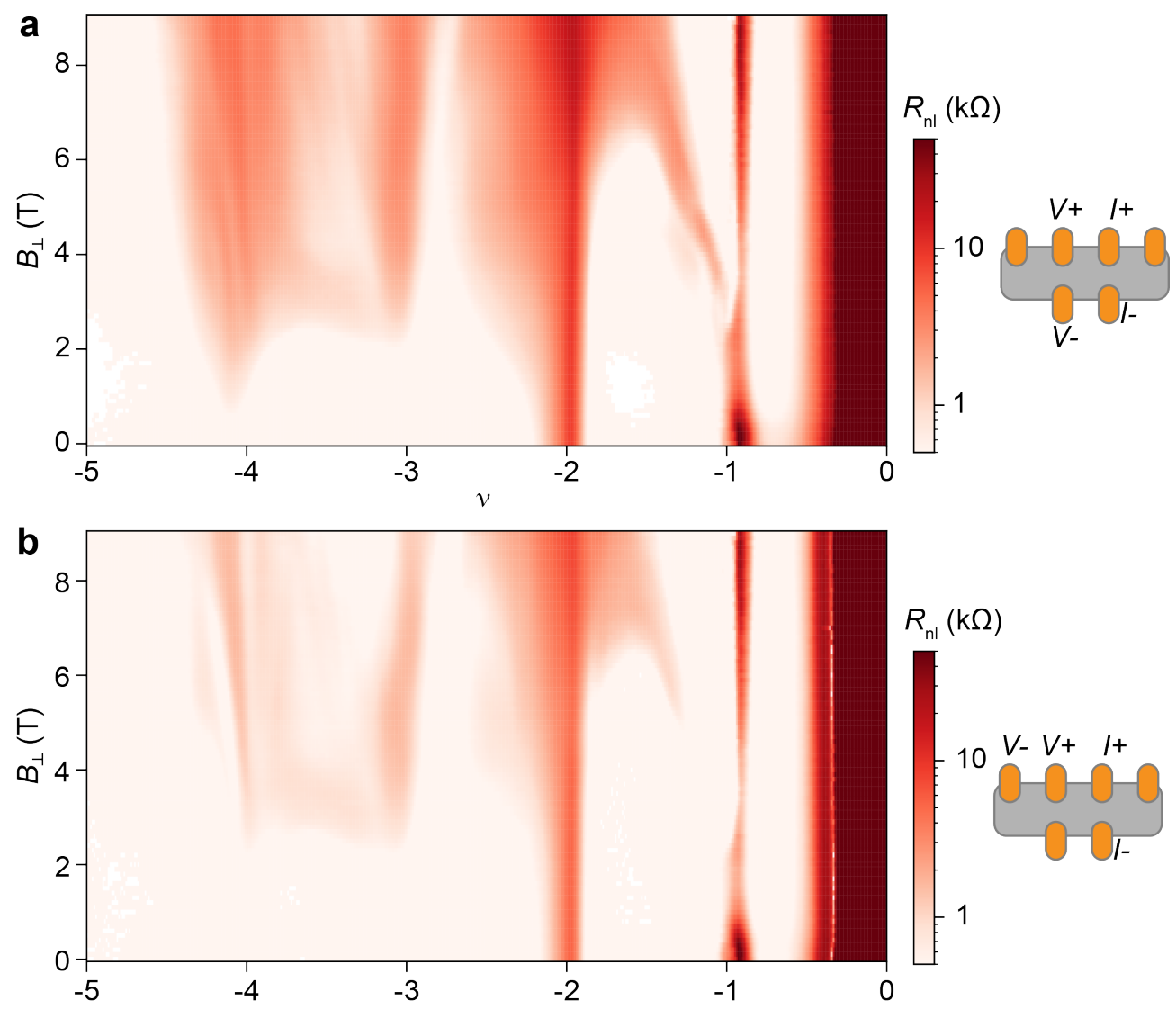

**Extended Data Figure 9| $R_{nl}$ as a function of $\nu$ and $B_\perp$ at $E = 0$ V/nm in two nonlocal measurement configurations (the electric connections are shown in the right panels).** Above a critical magnetic field of $B_\perp \approx 2.5$ T, a pronounced $R_{nl}$ emerges at $\nu = -3$, indicating the onset of long-range edge dominated transport. This behavior is consistent with the formation of a quantum spin Hall (QSH) state, where conduction is dominated by helical edge channels rather than bulk carriers.